\documentclass[12pt]{article}
\usepackage{amsmath,amssymb,epsfig}

\begin{document}

\begin{titlepage}
\rightline{February 2005}
\vskip 1.5 truecm
\begin{center}
{\Large\bf 
Quantum Field Theories and Critical Phenomena on Defects
}
\end{center}
\vskip 1.3cm
\centerline{Davide Fichera\footnote{fichera@sns.it}, 
Mihail Mintchev\footnote{mintchev@df.unipi.it}, 
Ettore Vicari\footnote{vicari@df.unipi.it}}

\vskip 0.4cm
\centerline{\sl Dipartimento di Fisica dell'Universit\`a di Pisa, Pisa, Italy}
\centerline {\sl Istituto Nazionale di Fisica Nucleare, Sezione di Pisa}
\centerline {\sl Largo Pontecorvo 3, 56127 Pisa, Italy}
\vskip 1.cm

\begin{abstract}

We construct and investigate quantum fields induced on a
$d$-dimen\-sional dissipationless defect by bulk fields propagating in a
$d+1$-dimensional space.  All interactions are localized on the
defect.  We derive a unitary non-canonical quantum field
theory on the defect, which is analyzed both in the continuum and on
the lattice. The universal critical behavior of the underlying system
is determined.  It turns out that the O($N$)-symmetric
$\varphi^4$ theory, induced on the defect by massless bulk fields,
belongs to the universality class of particular $d$-dimensional spin
models with long-range interactions.  On the other hand, in the
presence of bulk mass the critical behavior crossovers to the one of
$d$-dimensional spin models with short-range interactions.  

\vskip 1.cm

\noindent {\bf Keywords:}
Quantum Field Theories, Defects, Extra Dimensions, Critical Phenomena

\end{abstract}

\vfill
\rightline{IFUP-TH 06/2005}
\newpage
\pagestyle{plain}
\setcounter{page}{1}

\end{titlepage}

\section{Introduction}
\label{intro}

Canonical Lagrangian quantum field theory (QFT) dominates our present
understanding of elementary particle physics. It is well known
however, that in principle one can formulate consistent QFT's without
necessarily referring to a Lagrangian and the related canonical
formalism. Although may be less attractive in the context of
elementary particles, this possibility is relevant in the study of
critical phenomena. The enormous progress in conformal field theory
(CFT) in the past two decades has shown in fact (see e.g. \cite{CFT})
that only a small number of such models originate from a canonical
Lagrangian.

Inspired by the constant advance in using extra 
dimensions in QFT, we propose a new class of
non-canonical QFT's in $d$ space-time dimensions, induced by 
canonical fields in $d+1$ dimensions. The main steps of
our construction can be summarized as follows. One starts with a
conventional Lagrangian QFT with a local action $A[\Phi_i]$ for the
fields $\{\Phi_i\}$ defined on a $d+1$-dimensional manifold (bulk
space) ${\cal M}$. Then one considers a $d$-dimensional submanifold
${\cal D}\subset {\cal M}$, which can be interpreted physically as a
defect (impurity) in the bulk. We denote by $(x,y)$ the coordinates of
a generic point of ${\cal M}$ and assume that the defect ${\cal D}$ is
recovered for $y\to 0$. The bulk fields $\{\Phi_i(x,y)\}$ evolve
according to the Euler-Lagrange equations following from the bulk
action $A_{{}_{\cal M}}[\Phi_i]$, supplemented by initial conditions
in ${\cal M}$ and boundary conditions on ${\cal D}$.  The quantum
field $\varphi_i(x)$ induced by $\Phi_i(x,y)$ on ${\cal D}$ is
obtained by performing in appropriate way the limit $y\to 0$ in
$\Phi_i(x,y)$. In spite of the fact that $\{\varphi_i(x)\}$ are
non-canonical, their correlation functions define a meaningful local
QFT on the defect ${\cal D}$. This theory, which is unitary provided
that the defect does not dissipate, will be the goal of our
investigation in the present paper. The strategy summarized above
consists of two steps. One works first in the bulk, applying
standard techniques to the canonical action $A_{{}_{\cal
M}}[\Phi_i]$. Afterwards, one ``projects" the theory on the defect,
deriving an effective, non-canonical action $A_{{}_{\cal
D}}[\varphi_i]$ on ${\cal D}$.  An essential aspect of this framework
are the interactions, which are assumed to be localized on the defect
${\cal D}$. This idea has been already explored \cite{Saleur:1998hq}
in the context of two-dimensional conformal field theory, where a
specific exponential interaction localized at a point has been shown
to explain the edge states tunneling in the fractional quantum Hall
effect.

The paper is organized as follows. In section \ref{sec2} we
describe how free quantum fields are induced on defects. We discuss
the basic properties of the induced fields and derive the effective
action on the defect. In section 3 we investigate the effects of
interactions localized on the defect. In particular, we consider the
$d+1$-dimensional theory for an $N$-component scalar field $\Phi$ with
O($N$)-invariant $\Phi^4$ interaction localized on a $d$-dimensional
defect. We show that the theory can also be formulated in terms of a
$d$-dimensional O($N$) vector model defined on the defect and
interacting with a $d+1$-dimensional bulk free field.  Using general
renormalization-group arguments, we discuss the main features of the
universal critical behavior described by the unitary quantum field
theory induced on the defect.  In section 4 we solve the theory in the
large-$N$ limit, both in the continuum and on the lattice.  We
determine the critical behavior on the defect, and check the scenario
put forward in section 3. Finally, section 5 contains our conclusions.

\section{Free scalar field induced on a $\delta$-defect} 
\label{sec2}

In order to fix the ideas, we consider the simplest example of a bulk
scalar field $\Phi$ propagating in a $d+1$-dimensional Minkowski space
${\cal M}$, whose diagonal flat metric has signature $(+,-,...,-)$. We
adopt the coordinates $(x^0,...,x^{d-1},y) \in {\cal M}$ and study a
$\delta$-type impurity ${\cal D}$ localized on the $d$-dimensional
Minkowski space determined by $y=0$. The dynamics is defined by the
bulk action
\begin{equation}
A_{{}_{\cal M}}[\Phi] = -\frac{1}{2} \int_{-\infty}^\infty d^{d}x\, dy\,
\Phi(x,y)\left [\square_x - \partial_y^2 + M^2 + 2\mu \, \delta(y) \right
]\Phi(x,y) \, ,
\label{act} 
\end{equation} 
where $\mu$ characterizes the interaction of $\Phi$ with the defect. 
The variation of (\ref{act}) gives the 
equation of motion 
\begin{equation}
\left (\square_x - \partial_y^2 + M^2 \right ) \Phi (x,y) =  0 \, , 
\qquad y\not=0\, , 
\label{eqm}
\end{equation}
and the defect boundary conditions  
\begin{eqnarray}
&&\Phi(x,+0)=\Phi(x,-0) \equiv \Phi(x,0)\, , 
\label{bc1}\\
&&\partial_y\Phi (x,+0)-\partial_y\Phi(x,-0)=2\mu\, \Phi(x,0)\, . 
\label{bc2}
\end{eqnarray}
The quantum field $\Phi$, satisfying eqs. (\ref{eqm}-\ref{bc2}) and the
conventional equal-time commutation relations, is unique and is fully
determined by its two-point function.  Performing the quantization,
one must take into account that signals propagating in the bulk are
both reflected and transmitted \cite{Cherednik:jt, Delfino:1994nr} by the impurity. 
The relative reflection and transmission coefficients in momentum space read
\begin{equation}
R(p) = \frac{-i\mu}{p+i\mu}\, , \qquad T(p) =
\frac{p}{p+i\mu}
\label{RT}
\end{equation}
and satisfy unitarity
\begin{eqnarray}
T(p) T(-p) + R(p) R(-p) = 1\, , 
\label{unit1}\\
T(p) R(-p) + R(p) T(-p) = 0\, . 
\label{unit2}
\end{eqnarray}
These conditions imply the absence of dissipation on the
$\delta$-defect. The coefficients (\ref{RT}) deform the algebra of
canonical commutation relations to a reflection-transmission algebra
\cite{Mintchev:2003ue}, which is the main tool for quantizing
equations (\ref{eqm}-\ref{bc2}). Referring for the details to
\cite{Mintchev:2004jy}, we focus on the two-point vacuum expectation
value of $\Phi$. In the range $\mu \geq 0$ one gets 
\begin{equation} 
\langle \Phi (x_1,y_1)
\Phi(x_2,y_2)\rangle = \int_{-\infty}^{\infty} \frac{dp}{2\pi}
E(p;y_1,y_2;\mu) W_{M^2+p^2}(x_{12})\, .  
\label{2P}
\end{equation}
Here $x_{12}\equiv x_1-x_2$, 
\begin{eqnarray}
&&E(p;y_1,y_2;\gamma) = 
\theta (y_1)\theta (-y_2) T(p) e^{ipy_{12}} + 
\theta (-y_1)\theta (y_2) T(-p) e^{ipy_{12}} +
\label{E}
\\
&&\;\;\theta (y_1)\theta (y_2)\left [e^{ipy_{12}} + 
R(p) e^{ip{\widetilde y}_{12}}\right ] 
+\theta (-y_1)\theta (-y_2)\left [e^{ipy_{12}} + 
R(-p) e^{ip{\widetilde y}_{12}}\right]   
\nonumber 
\end{eqnarray}
with ${\widetilde y}_{12} = y_1+y_2$ and 
\begin{equation} 
W_{m^2}(x)=\int_{-\infty}^{\infty} d_{d}k 
e^{-ikx} \theta (k^0)\, 2\pi \delta (k^2-m^2) \, , \qquad 
d_nk \equiv \frac{d^nk}{(2\pi)^n}\, , 
\label{W}
\end{equation}
is the standard two-point vacuum expectation value of a free 
scalar field of mass $m$ in $d$ space-time dimensions. 

The quantum field $\varphi (x)$, induced on the $\delta$-defect, 
is defined by the weak limit 
\begin{equation} 
\varphi (x) = \sqrt {2} \lim_{y \to 0} \Phi (x,y)  
\label{lim}
\end{equation}
in the Hilbert space of $\Phi (x,y)$. The factor $\sqrt 2$ is 
introduced for further convenience. The 
limit ({\ref{lim}) exists \cite{Mintchev:2004jy} and 
after a straightforward change of variables in (\ref{2P}) one gets 
\begin{equation} 
\langle \varphi (x_1) \varphi(x_2)\rangle = 
\int_{M^2}^{\infty} \frac{d\lambda^2}{\pi} 
\frac{\sqrt {\lambda^2 - M^2}}{\lambda^2 - M^2 +\mu^2} 
W_{\lambda^2}(x_{12})\, .   
\label{2}
\end{equation}
Eq. (\ref{2}) is a K\"all\'en-Lehmann spectral 
representation with density 
\begin{equation} 
\varrho (\lambda^2) = \theta(\lambda^2 - M^2) 
\frac{\sqrt {\lambda^2 - M^2}}{\pi(\lambda^2 - M^2 +\mu^2)} \, ,  
\label{rho}
\end{equation}
which gives rise to a well-defined generalized free field \cite{J} on ${\cal D}$. 
In the context of QFT with extra dimensions 
the induced field $\varphi$ is a superposition of Kaluza-Klein (KK) 
modes with mass $\lambda$. The positive function $\varrho$ defines 
the KK measure. Since this measure is polynomially bounded at infinity, 
$\varphi $ is a {\it local} quantum field on the defect.\footnote{The 
locality properties of induced quantum fields are investigated 
in \cite{Mintchev:2001mh}.} Obviously it does not satisfy equal-time canonical
commutation relations and is therefore non-canonical. Moreover, $\varphi$ has 
a continuum mass spectrum with mass gap $M$. 

{}From (\ref{2}) one obtains the propagator 
\begin{equation} 
\tau(x_{12}) \equiv \langle T\varphi (x_1) \varphi(x_2)\rangle = 
\int_0^\infty d\lambda^2 \varrho(\lambda^2)\, \Delta_{\lambda^2}(x_{12}) \, ,
\label{dprop}
\end{equation} 
where 
\begin{equation}
\Delta_{m^2}(x) = \frac{1}{i}\int_{-\infty}^\infty 
d_{d}k \, \frac{e^{-ikx}}{m^2 - k^2 - i\varepsilon} \, , 
\label{prop}
\end{equation} 
is the familiar propagator in $d$ space-time dimensions. 
The integral in (\ref{dprop}) is easily computed 
and gives for the Fourier transform ${\widehat \tau}$ of the 
propagator the result 
\begin{equation} 
{\widehat \tau}(k) = \frac{1}{i}
\frac{1}{\sqrt{M^2-k^2-i\varepsilon}+\mu}\, .  
\label{pdprop}
\end{equation}
It is worth stressing that the poles, that are 
present in the single KK modes in the complex
$k_0$-plane, give rise to a cut after the re-summation. 

For Euclidean momenta $(k_1,...,k_d)$ one gets from
(\ref{pdprop}) the Schwinger function  
\begin{equation} 
{\widehat s}(k_1,...,k_d) = 
i\, {\widehat \tau}(k_0 = -ik_d,k_1,...,k_{d-1}) = 
\frac{1}{\sqrt{M^2+k^2}+\mu} \, ,    
\label{mpropagator}
\end{equation} 
which leads to the following Euclidean effective action 
\begin{equation}
A_{{}_{\cal D}}[\varphi] = \frac{1}{2} \int_{-\infty}^\infty d^{d}x \, 
\varphi (x) \left (\sqrt{M^2 -\Delta} + \mu \right ) \varphi (x) \, ,  
\label{act1} 
\end{equation}
localized on the defect. Eq. (\ref{act1}) reads in momentum space 
\begin{eqnarray}
&&A_{{}_{\cal D}}[\varphi] = \frac{1}{2} \int_{-\infty}^\infty d_{d}k \, 
{\widehat \varphi} (-k) J(k;M,\mu)\,  
{\widehat \varphi} (k) \, , \label{act1k} \\ 
&&J(k;M,\mu)\equiv \sqrt{k^2+M^2} + \mu \, , 
\nonumber
\end{eqnarray}
where  
\begin{equation} 
\int_{-\infty}^\infty d_{d}k e^{-ikx} J(k;0,0) \sim \frac{1}{r^{d+1}} \, , 
\qquad r\equiv |x|\, .  
\label{pot}
\end{equation}
In what follows eq. (\ref{pot}) allows to make contact with some previous 
work \cite{FMN-72}-\cite{Suzuki-73} in the context of 
spin models in statistical mechanics. 

The case $-M\leq \mu <0$ can be considered along the same lines, keeping in mind that 
the interaction of $\Phi$ with the defect produces \cite{Mintchev:2004jy} a defect bound state. 
It contributes to the Schwinger function, which instead of (\ref{mpropagator}), 
now takes the form \cite{Mintchev:2000mf}
\begin{equation} 
{\widehat s}(k) = \frac{1}{\sqrt{M^2+k^2}+\mu} + 
\theta(-\mu) \frac{2|\mu|}{k^2+M^2-\mu^2}\, .     
\label{mpropagatorbstate}
\end{equation} 
In the range $\mu <-M$ the defect boundary state gives imaginary 
energy contribution to $\Phi$, which leads to an instability. 
Therefore the free theory (\ref{act}) is stable and well-defined only when $\mu >-M$. 

Let us mention in conclusion that the $\delta$-impurity is actually an
element of a whole three-parameter family \cite{Albeverio} of
dissipationless defects, defined by the boundary conditions
\begin{equation}
\left(\begin{array}{cc} \varphi (x,+0) \\ 
\partial_y \varphi (x,+0)\end{array}\right) = 
\left(\begin{array}{cc} a_{11} & a_{12}\\ a_{21} & a_{22}\end{array}\right)
\left(\begin{array}{cc} \varphi (x,-0) \\ 
\partial_y \varphi (x,-0)\end{array}\right)\, ,  
\label{bc}
\end{equation} 
where $\{a_{ij} \in {\mathbb R} \, :\, a_{11}a_{22}-a_{12}a_{21} = 1\}$. The above
considerations have a straightforward extension \cite{Mintchev:2004jy}
to all of them. For the K\"all\'en-Lehmann spectral density one gets in 
the general case 
\begin{equation} 
\varrho (\lambda^2) = \theta(\lambda^2 - M^2) 
\frac{2\left [a_{12}^2(\lambda^2 - M^2) +a_{11}^2+1\right]\sqrt {\lambda^2 - M^2}}
{\pi\left[a_{12}^2(\lambda^2 - M^2)^2 +(a_{11}^2+a_{22}^2+2)
(\lambda^2 -M^2)+a_{21}^2\right ]} \, ,   
\label{rhogeneral}
\end{equation}
which behaves for $\lambda=0$ and $\lambda = \infty$ like 
the density (\ref{rho}). For this reason it is enough to concentrate 
in what follows on the $\delta$-impurity. Most of our results 
hold in fact in the general case, because they reflect the low-momentum behavior 
of the field $\varphi$.

\section{$\Phi^4$ interactions on $\delta$-defects} 
\label{sec3}

In this section we extend the study to the case in which $\Phi^4$-type
interactions are present on the defect and the dynamics is determined
by the euclidean bulk action
\begin{equation}
A_{{}_{\cal M}}[\Phi] = 
\int_{-\infty}^{\infty} d^d x dy\left [{1\over 2} (\partial\Phi)^2
+ {M^2\over 2} \Phi^2 +
\delta(y)\, \mu\, \Phi^2 + \delta(y)\, {g_0\over 4!} (\Phi^2)^2 \right]\, , 
\label{acphi}
\end{equation}
where $\Phi$ is an $N$-component scalar field.  We should mention that
surface (boundary) critical phenomena have been widely investigated in
the case the $\Phi^4$-type potential determines also the critical properties in
the bulk, see for example the reviews \cite{Binder-83, Diehl-86}. Here, the
defect does not only represent the impurity of the bulk system, but it
is also the place where the $\Phi^4$ interaction is localized. 
As already mention in the introduction, the idea of localizing 
the interaction on the defect was already considered \cite{Saleur:1998hq} 
in the context of two-dimensional conformal field theory and the 
quantum Hall effect. In the following we discuss the phase
diagram and the critical properties on the $d$-dimensional defect $\cal D$,
with $d\ge 2$, and especially how they are affected by the presence of
a free field in the bulk.

This theory can be generalized by introducing a field
$\varphi$ defined on the defect, and considering the action
\begin{eqnarray}
A[\varphi, \Phi] = &&\int_{-\infty}^\infty d^d x 
\left[ \frac{\kappa}{2}(\partial\varphi)^2(x) + \mu \varphi^2(x) + 
{g_0\over 4!}\varphi^4(x) + \frac{h}{2}\left [\varphi(x)-\Phi(x,0)\right ]^2 \right] 
\nonumber \\
+ &&\int_{-\infty}^\infty  d^d x dy \left[ {1\over 2} (\partial \Phi)^2(x,y) + 
{M^2\over 2} \Phi^2(x,y)\right]\, ,  
\label{acpsivarphi}
\end{eqnarray}
where $\kappa \geq 0$ and $h>0$ are real parameters. 
One can easily check that in the limit $\kappa \rightarrow 0$ and 
$h\rightarrow \infty$ the theory (\ref{acphi}) is recovered (apart from a trivial
rescaling of the field). In this model the field $\varphi$, constrained
on a $d$-dimensional defect, interacts with a free field $\Phi$
propagating in $d+1$ dimensions.  Moreover, as in the case of the
standard O($N$)-symmetric $\varphi^4$ theories, one may also consider a particular
limit of the parameters $\mu$ and $g_0$ so that the resulting
defect field $\varphi$ is constrained to have norm one, i.e. 
\begin{eqnarray}
A[\varphi, \Phi] = &&\int_{-\infty}^\infty d^d x 
\left[ \frac{\kappa}{2}(\partial\varphi)^2(x) + \frac{h}{2}\left
[\varphi(x)-\Phi(x,0)\right ]^2 \right] 
\nonumber \\
&&+ \int_{-\infty}^\infty  d^d x dy \left[ {1\over 2} (\partial \Phi)^2(x,y) + 
{M^2\over 2} \Phi^2(x,y)\right]\, ,  
\label{acsigvarphi}
\end{eqnarray}
where $\varphi \cdot \varphi \equiv \sum_{i=1}^N\varphi_i\varphi_i=1$.
Note that in the limit $h\rightarrow \infty$, this model can be seen
as a free field $\Phi$ constrained to have norm one on the defect.
The universal critical behavior is expected to remain unchanged under
the above changes of the Lagrangian parameters.  This can be verified
in the large-$N$ limit, see section~\ref{sec4}.  In particular, the
parameters $\kappa$ and $h$ are expected to be irrelevant from the
point of view of the renormalization-group theory, since they do not
change the universal behavior at its critical points. They may be set
to the values $\kappa=0$ and $h=\infty$.

The above models can be regularized on a $d+1$-dimensional lattice by a
straightforward discretization of their Lagrangians.  For example, 
a straightforward discretization of the action (\ref{acsigvarphi}) is
\begin{equation}
A_{{}_{\cal L}} = - \kappa \sum_{x,\mu} \varphi_x \varphi_{x+\mu} 
+ \frac{h}{2} \sum_x (\varphi_x - \Phi_x)^2 +
\frac{1}{2} \sum_{z,\nu} (\Phi_z - \Phi_{z+\nu})^2
+ \frac{M^2}{2} \sum_z \Phi_z^2 \, , 
\label{aclat}
\end{equation}
where we set the lattice spacing $a=1$, 
$x$ and $z$ indicate respectively the coordinates on the
defect and in the bulk, and $\mu=1,...,d$, $\nu=1,...,d+1$.
The corresponding partition function is defined as
\begin{equation}
Z_{{}_{\cal L}} = \sum_{\{\Phi\},\{\varphi\}} 
\prod_x \delta(\varphi_x^2-1) e^{-\beta A_{{}_{\cal L}}} \, , 
\label{pafu}
\end{equation}
where $\beta=1/T$, and the coupling $T$ plays the role of temperature.
Nontrivial continuum limits are realized at the critical points, where
a continuous transitions occur and a length scale diverges in unit of
the lattice spacing.

The main features of the phase diagram can be discussed using general
renormali\-zation-group arguments.  In the case $h=0$, i.e. when the
fields $\varphi$ and $\Phi$ do not interact, the critical properties
of the $\varphi$ field on the defect are those of the $d$-dimensional
$N$-component nonlinear sigma model.  If $d=1$ the correlation length
diverges only when $T\rightarrow 0$ for any $N$.  If $d=2$, the system
undergoes a finite-temperature Ising transition for $N=1$, a
finite-$T$ Kosterlitz-Thouless transition \cite{KT-73} for $N=2$,
while in the case $N\ge 3$ the system becomes critical only in the
limit $T\rightarrow 0$, with a length scale $\xi$ that increases
exponentially, i.e. $\xi\sim e^{b/T}$, typical of asymptotically free
models.  For $d=3$ there is a finite-temperature transition for any
$N$ with ujniversal properties characterized by nontrivial power
laws.~\footnote{The universal properties of three-dimensional
 O($N$) vector models
have been determined by various theoretical methods and
experiments. The most precise theoretical estimates of the standard
critical exponents have been obtained by lattice techniques.  We
mention $\nu=0.63012(16)$, $\eta=0.03639(15)$ \cite{CPRV-02} and
$\nu=0.63020(12)$, $\eta=0.0368(2)$ \cite{DB-03} for the 
three-dimensional Ising model
corresponding to $N=1$, $\nu=0.67155(27)$ and $\eta=0.0380(4)$
\cite{CHPRV-01} for the XY universality class ($N=2$), and
$\nu=0.7112(5)$ and $\eta=0.0375(5)$ \cite{CHPRV-02} for the Heisenberg
universality class ($N=3$). In the large-$N$ limit one finds
$\nu=1$ and $\eta=0$ \cite{ZJbook}.}  We have the same scenario for $d=4$ but
with mean-field critical behaviors apart from logarithms.  For $d>4$
the behavior is just mean field.  See, e.g., refs.~\cite{ZJbook,PV-02}
for reviews.  These critical behaviors remain unchanged for $h>0$ when
$M>0$, i.e. when $M$ is strictly positive and kept fixed.  In this
case the large-distance correlations in the bulk decay exponentially,
i.e. $\sim e^{-M r}$, inducing only short-ranged interactions on the
defect. They do not change the universal critical behavior of the low
modes when the correlation length $\xi$ on the defect is sufficiently
large, i.e. when $\xi\gg 1/M$.  On the other hand, when the bulk field
$\Phi$ becomes massless, the large-distance bulk correlations induce
long-range interactions on the defect, cf. eq.~(\ref{pot}), which can
change the critical properties on the defect.  These long-range
interactions can be inferred from the analysis of section~\ref{sec2}.
When $M=0$, integrating out the bulk field $\Phi$ in
eqs.~(\ref{acsigvarphi}) and (\ref{aclat}), we expect to obtain a
defect action of the type
\begin{eqnarray}
&&A_{{}_{\cal D}} = \int_{-\infty}^\infty 
d_dk\, J(k)\, {\widehat \varphi} (-k) \cdot {\widehat \varphi} (k) \, ,  
\label{defh} \\
&&J(k)\simeq j_{1} (k^2)^{1/2} + j_2 k^2 + ... \, .
\nonumber
\end{eqnarray}
The critical properties of statistical
systems with long-range interactions, such as
\begin{equation}
J(k) = j_s (k^2)^{s/2} + j_2 k^2 + ...
\label{jks}
\end{equation}
were studied in \cite{FMN-72,Sak-73} within an expansion in powers of
$\epsilon \equiv 2s-d$ and in \cite{Suzuki-73} in powers of $1/N$.
The results of \cite{FMN-72}--\cite{Suzuki-73} show that statistical
systems with Hamiltonians of the type (\ref{defh}) undergo a
continuous transition at finite temperature for any $N$. The cases
$d>2$, corresponding to $\epsilon<0$, are in the classical regime
\cite{FMN-72}, where, for any $N$, the critical behavior of the
magnetic susceptibility $\chi$ and correlation length $\xi$ are given
by \cite{FMN-72}
\begin{equation}
\chi \sim \xi \sim t^{-1}\, , 
\label{clcb}
\end{equation}
where $t\equiv (T-T_c)/T_c$ is the reduced temperature;
therefore the critical exponents are $\gamma=\nu=\eta=1$.  The case
$d=2$ is on the borderline $\epsilon=0$, and
multiplicative logarithms correct the classical behavior,
\cite{FMN-72}
\begin{equation}
\chi \sim \xi \sim t^{-1} ( \ln t^{-1} )^{N+2\over N+8}\, .
\label{clcbd2}
\end{equation}
We also mention that for $d<2$,
corresponding to $\epsilon>0$, the critical exponents are nontrivial,
indeed
\begin{equation}
\nu= 1 + \frac{N+2}{N+8}
\epsilon + O(\epsilon^2)\, ,\qquad
\eta=1 + O(\epsilon^2)\, .
\label{expeps}
\end{equation}
The $O(\epsilon^2)$ terms are reported in refs.~\cite{FMN-72,Sak-73}. 

The above-mentioned critical properties suggest that the critical point
for $M=0$ is actually a tricritical point in the $T$-$M$
plane.  Indeed, beside the two standard relevant scaling fields
associated with the temperature $T$ and external field $H$, there is
another relevant parameter given by the bulk mass $M$.  Switching $M$
on, the critical behavior crossovers to the more stable one for $M>0$.
We will return to this point in section \ref{sec4}.

We finally note that the critical properties discussed in this section
apply also to the more general dissipationless defects defined by 
eq. (\ref{bc}), since the low-momentum behavior of
the corresponding effective Lagrangian remains substantially unchanged.

\section{The large-$N$ limit} 
\label{sec4}

In this section we show how the scenario put forward in the preceding
section is actually realized in the large-$N$ limit, which can be
analytically investigated by solving the corresponding saddle-point
equations. We first discuss the large-$N$ limit in an appropriate 
continuum formulation. 
Then we consider a more rigorous non-perturbative treatment of 
the bulk theory based on a lattice
regularization of the path integral. In other words, we 
consider a lattice model whose critical
behavior is described by the unitary quantum field 
theory induced on the defect. We
present the solution of its large-$N$ limit for $d=2$.  As we shall
see, this large-$N$ investigation fully supports the scenario outlined 
in section \ref{sec3}.

\subsection{Continuum theory}
\label{sec4s1}

We want to study the effects of the $\Phi^4$ interaction localized on 
the defect, and in particular the universal properties near the critical point 
at which the bulk correlation lenght diverges, i.e. for $M\rightarrow 0$. For this purpose, 
the results of section~\ref{sec2} suggest to introduce 
the following effective model defined on the defect by  
\begin{equation} 
A_{{}_{\cal D}}[\varphi] = \int_{-\infty}^\infty
d^{d}x \left [ \frac{1}{2}\sum_{i=1}^N\varphi_i \left (\sqrt{M^2
-\Delta} + \mu \right ) \varphi_i + \frac{g_0}{4!}\left
(\sum_{i=1}^N\varphi_i \varphi_i\right )^2 \right ]\, .
\label{act2} 
\end{equation}
Note that the above theory is renormalizable
in two dimensions, unlike the standard $\varphi^4$ model 
which is renormalizable in four dimensions. 

In the following we first determine the critical behavior of the 
theory (\ref{act2}) at $M=0$, computing in particular the standard critical
exponents and the equation of state in the limit $N\to \infty$ with
fixed $g_0N\equiv g$. We introduce the auxiliary field
$\lambda$ and a source $H$ for $\varphi_{{}_N}$, writing the partition
function in the form
\begin{equation} 
Z(H) = \int [d\lambda][d\varphi] e^{-A_{{}_{\cal D}}[\varphi, \lambda;H]}\, , 
\label{pf1}
\end{equation}
where 
\begin{equation}
A_{{}_{\cal D}}[\varphi, \lambda;H] = \int_{-\infty}^\infty d^{d}x 
\left [ \frac{1}{2} \sum_{i=1}^N\varphi_i 
\left (\sqrt{-\Delta} + \lambda \right ) \varphi_i 
 + \frac{3\mu }{g_0}\lambda - \frac{3}{2g_0}\lambda^2 - H\varphi_{{}_N} \right ]\, .  
\label{act3} 
\end{equation} 
Integrating over $\varphi_1,...,\varphi_{{}_{N-1}}$ and setting 
$\sigma = \varphi_{{}_N}$, one gets 
\begin{equation} 
Z(H) = \int [d\lambda][d\sigma] e^{-A_{{}_{\cal D}}[\sigma, \lambda;H]}\, , 
\label{pf2}
\end{equation}
where 
\begin{eqnarray} 
A_{{}_{\cal D}}[\sigma, \lambda;H]= 
\int_{-\infty}^\infty d^{d}x 
\Biggl [ \frac{1}{2}\sigma 
\left (\sqrt{-\Delta} + \lambda \right ) \sigma + \quad  
\nonumber \\ 
\frac{3\mu }{g}\lambda - \frac{3}{2g}\lambda^2  
+ \frac{N-1}{2}\, {\rm Tr} \log \left (\sqrt{-\Delta }+ \lambda \right ) 
- H\sigma \Biggr ]\, .  
\label{act4}
\end{eqnarray}
The large-$N$ behavior is governed by the uniform ($\sigma =$ const.,  
$\lambda =$ const.) saddle-point approximation. Varying (\ref{act4}) 
and rescaling $\sigma $ and $H$ according to $\sigma \mapsto \sigma \sqrt{N}$ 
$H \mapsto H \sqrt{N}$, one gets
\begin{eqnarray} 
&&\lambda \sigma = H  \,,
\label{sp1} \\
&&\sigma^2 - \frac{6}{g} (\lambda - \mu) + 
\int^\Lambda d_dk \frac{1}{|k|+\lambda } = 0 \, . 
\label{sp2}
\end{eqnarray}
The integral in (\ref{sp2}) is regularized by means of an UV cutoff
$\Lambda$. Moreover, we take $d>1$ in order to avoid IR instabilities.
As expected, eqs. (\ref{sp1},\ref{sp2}) strongly resemble their
counterparts \cite{ZJbook} in the conventional $\varphi^4$ theory.
The only difference is the term $|k|$ in the denominator of the
integrand, which replaces $k^2$ from the standard $\varphi^4$ model.
The consequences of this modification are easily analyzed.  

Let us consider first the case $H=0$.
In the broken phase $\sigma\not=0$, one has $\lambda
=0$. Eq. (\ref{sp2}) has a solution for $\sigma $ only if
\begin{equation} 
\mu < \mu_c \equiv 
-\frac{g}{6}\int^\Lambda d_dk \frac{1}{|k|} \, . 
\label{muc}
\end{equation} 
Setting 
\begin{equation}
t \equiv \frac{6}{g}(\mu - \mu_c)\, , 
\label{tdef}
\end{equation} 
one gets from (\ref{sp2})  
\begin{equation} 
\sigma^2 = -t \sim (-t)^{2\beta} \, ,
\label{beta} 
\end{equation} 
which implies $\beta = \frac{1}{2}$.  

In the symmetric phase $\sigma=0$ and $\lambda \equiv m\not=0$, where  
$m$ is the inverse correlation length $\xi$, which determines the large-distance 
exponential decay of the two-point correlation function of
the $\varphi$ field, i.e. $G(r)\sim e^{-r/\xi}$. Now eq. (\ref{sp2})
takes the form
\begin{equation} 
\frac{6}{g} + \int^\Lambda d_dk \frac{1}{|k|(|k|+m)} = 
\frac{t}{m} \, .
\label{sp3}
\end{equation}
{}From (\ref{sp3}) one deduces 
\begin{equation} 
m \sim t^\nu \, , \qquad \nu = 
\left\{\begin{array}{cc}
1 \, , & \quad \mbox{$d>2$}\, ,\\[1ex]
\frac{1}{d-1}\, ,
& \quad \mbox{$1<d<2$}\, .\\[1ex]
\end{array} \right.
\label{nu}
\end{equation}
{}From the critical behavior of
the two-point correlation function one finds $\eta = 1$. These 
results agree with the formulas (\ref{expeps}) 
obtained within an $\epsilon\equiv 2-d$ expansion.
The case $d=2$, which is special because the integral in (\ref{sp2})
generates the logarithm $\ln\frac{\Lambda}{m}$, will be considered 
in the next subsection on the lattice. One has mean-field
behavior with logarithmic corrections. 

We turn now to the case $H\not=0$. Combining (\ref{sp1}) and 
(\ref{sp2}) one gets 
\begin{equation} 
\sigma^2 + t = 
\frac{6}{g}m + m\int^\Lambda d_dk \frac{1}{|k|(|k|+m)} \, . 
\label{h}
\end{equation} 
In the domain $1<d<2$ it implies
\begin{eqnarray}
&&H \sim \sigma^\delta f(t \sigma^{-\frac{1}{\beta}}),\qquad
f(x) = ( 1 + x )^\gamma,
\label{scaleq}\\
&&\delta = \frac{d+1}{d-1}\, , 
\qquad \gamma = \frac{1}{d-1} ,
\label{deltagamma}
\end{eqnarray}
where $f(x)$ is a universal function apart from trivial normalizations
(usually one sets $f(0)=1$ and $f(-1)=0$~\cite{ZJbook}).  
Note that the critical exponents satisfy the scaling and hyperscaling relations
\begin{equation} 
\beta = \frac{\nu}{2} (d-2+\eta)\, , \qquad 
\gamma = \nu (2-\eta)\, , \qquad 
\delta = \frac{d+2-\eta}{d-2+\eta}\, ,
\label{rel}
\end{equation}  
in the range $1<d<2$.  
{}From eq.~(\ref{scaleq}) one may also derive the effective potential
(Helmholtz free energy) $V(z)$, i.e. the generator of one-particle
irreducible correlation functions of $\varphi$ at zero momentum,
see, e.g., refs.~\cite{PV-02,PV-98}. We obtain
\begin{equation}
V(z) = \frac{3}{d} 
\left[ \left( 1 + \frac{d-1}{6} z^2 \right)^{d/(d-1)} - 1\right]\, , 
\label{effpotln}
\end{equation}
where $z\sim \sigma t^{-\beta}$, which represents the
renormalized expectation value of the field $\varphi$. In
eq.~(\ref{effpotln}) $V(z)$ has been normalized according to
$V(z)=\frac{1}{2} z^2 + \frac{1}{24} z^4 + O(z^6)$.
For $d=2$ the effective potential reduces to the simple form
\begin{equation}
V(z) = \frac{1}{2} z^2 + \frac{1}{24} z^4\, ,
\end{equation}
as in the case of the standard $\varphi^4$ model in four dimensions
\cite{PV-98}.

It is worth noting that the theory (\ref{act2}) induced on a
$d$-dimensional defect with $1<d<2$ presents analogies with the
standard $\varphi^4$ model for $2<d<4$, see, e.g., ref.~\cite{ZJbook}.
In both cases we have critical behaviors characterized by nontrivial
power laws. Moreover, replacing $d$ with $d/2$ in eqs.~(\ref{scaleq}),
(\ref{deltagamma}) and (\ref{effpotln}), one recovers the
corresponding expressions for the large-$N$ limit of the standard
O($N$) vector models.

As already discussed in section~\ref{sec3}, the critical behavior found
for $M=0$ is unstable against the the parameter $M$, since in the
presence of a bulk mass the $M=0$ critical behavior is not anymore
observed. In this case, the models are expected to show the
critical behavior of the well known $d$-dimensional O($N$) vector
universality class with short-range interactions, i.e. when
$J(k)\simeq k^2$ in eq.~(\ref{defh}).  Therefore, the critical point
for $M=0$, $T=T_{tc}$, is a tricritical point in the $T$-$M$ plane.
Renormalization-group arguments applied to tricritical points,
see, e.g., the review \cite{LS-84}, suggest the following
generalized scaling hypothesis (for $H=0$)
\begin{equation}
F_{\rm sing} \approx t^{d\nu} f\left(M t^{-\phi}\right),
\qquad \chi \approx t^{-\gamma} f_\chi\left(M t^{-\phi}\right),
\label{scaling}
\end{equation}
where $F_{\rm sing}$ is the singular part of the free energy, $\chi$
is the defect magnetic susceptibility; $t\equiv (T-T_{tc})/T_{tc}$,
$\nu$, $\gamma$, $\phi$ are the tricritical exponents. In particular
$\phi$ is the crossover exponent associated with the instability
direction parametrized by $M$. The scaling behavior (\ref{scaling})
is expected to hold in the critical crossover limit $t,M\rightarrow 0$
keeping $M t^{-\phi}$ fixed. This issue can be investigated
in the large-$N$ limit, starting from the action (\ref{act2}),
as we did in this section, but keeping $M\not=0$ through the various steps
to arrive at the saddle-point equation. Setting $H=0$, we obtain
\begin{equation}
\frac{6}{g} (M - m + \mu) + 
\int^\Lambda d_dk \frac{1}{m -M +\sqrt{M^2+k^2}} = 0\, , 
\label{eqM}
\end{equation}
where now $m$ is related to the zero-momentum correlation function $\chi $ (usually called 
magnetic susceptibility) by $\chi = 1/m$. Defining $t$ by eq.~(\ref{tdef}), one finds 
\begin{equation}
t  = {6 \over g} (M+m) +
\int_{-\infty}^{\infty} d_dk\, \,  
\frac {m -M +\sqrt{M^2+k^2}- \sqrt{k^2}}{\sqrt{k^2}\, (m -M + \sqrt{M^2+k^2})}\, . 
\label{eqM2}
\end{equation}
Keeping only the leading terms in the critical crossover limit for
$1<d<2$, we find
\begin{eqnarray}
&& t \approx m^{d-1} S(M/m),  \label{eqM3} \\
&& S(r) = \int_0^{\infty} d\rho \rho^{(d-2)}\, \, 
\frac{1 -r + \sqrt{r^2+\rho^2}-\rho} 
{1-r+\sqrt{r^2+\rho^2}}\, .  
\nonumber
\end{eqnarray}
This result is in agreement with the expected scaling (\ref{scaling}),
and implies $\phi=1/(d-1)$.  We also mention that in the classical
regime $d>2$, one finds $\phi=1$, while for $d=2$ the classical
crossover relations must be corrected due to multiplicative
logarithms, as in eq.~(\ref{clcbd2}). It is worth mentioning that the 
relation between $\chi$ and $\xi$ in the crossover regime changes: 
one has $\chi \sim \xi$ in the limiting case $M/m=0$, while in the opposite limit $M/m=\infty$, 
$\chi \sim \xi^2$.

\subsection{Theory on the lattice}
\label{sec4s2}

In this section we study the large-$N$ limit of the lattice model
defined by the action (\ref{aclat}). We consider lattices of
size $L^d\times E$ with periodic boundary conditions along all 
$d+1$ directions. All dimensional parameters are expressed below in 
units of the  lattice spacing $a$, which is set to 1 for convenience. 
To begin with, we integrate out the bulk field $\Phi$
in the partition function (\ref{pafu}), obtaining a $d$-dimensional
nonlinear sigma model defined on the defect.  Straightforward
calculations, based on Gaussian integrals, show that its effective
action in momentum space reads  
\begin{equation}
A_{\rm eff}= \frac{1}{2} 
\sum_k  [ \kappa \hat{k}^2 - B(k;M,h,E) ] \varphi_{-k} \cdot \varphi_k \, , 
\label{leff}
\end{equation}
where $\hat{k}^2= 2 \sum_\mu (1 - \cos k_\mu)$, $k_\mu=2\pi n_\mu/L$
with $n_\mu=1,...,L$. The function $B$, following from the integration, is
formally given by 
\begin{equation}
B(k;M,h,E) = h^2 Q^{-1}_{1,1}\, , 
\label{q11}
\end{equation}
where $Q$ is the $E\times E$ matrix
\begin{equation}
  Q = (M^2+\hat{k}^2)\mathbb{I} + \left(
  \begin{array}{ccccc}
  2 & -1 & 0 & \cdots & -1 \\
  -1&\ddots&\ddots&\ddots& \\
  0& \ddots& & & \\
  \ddots& & & &-1 \\
  -1&0&\cdots &-1&2
  \end{array}
  \right)
  +
  h
  \left(
  \begin{array}{cccc}
  1&0&0&\cdots \\
  0&0&\cdots & \\
  \vdots & & & \end{array}
  \right)\, . 
\label{QQ}
\end{equation}
We write the partition function as
\begin{equation}
Z_{\cal L} = \sum_{\{\varphi\}} 
\prod_x \delta(\varphi_x^2-1) e^{-\beta N A_{\rm eff}}\, . 
\label{pafu2}
\end{equation}

In the large-$N$ limit keeping $\beta$ fixed, the solution is given
by the saddle-point equation, which, in the limit $L\rightarrow \infty$, 
takes the form
\begin{equation}
  \beta=\int_{-\pi}^{\pi} d_d k \,
\frac{1}{m + \kappa \hat{k}^2 + [B(0;M,h,E)-B(k;M,h,E)]}\, .
\label{gapeq}
\end{equation}
Here $m$ is related to the the magnetic susceptibility $\chi$ by
$m=(\beta\chi)^{-1}$.  In the large-$N$ limit,
$\chi\sim \xi$ for $M=0$, while $\chi\sim\xi^2$ for $M>0$.
According to eq.~(\ref{q11}), in order to get the function $B$,
we must evaluate the 1,1 matrix element
of the inverse matrix of $Q$ given in Eq.~(\ref{QQ}).
In the limit $E\rightarrow \infty$ one finds
\begin{equation}
B(k;M,h,\infty)= - 
  \frac{h^2}{h + [ 4 M^2 + 4 \hat{k}^2
  + (M^2 + \hat{k}^2)^2 ]^{1/2} }\, . 
\label{bbf}
\end{equation}
This result is derived in the appendix.

In the following we simplify the calculation by fixing $\kappa=0$ and
$h=\infty$, but we have also checked that the universal critical
behavior remains the same when choosing generic values $\kappa,h>0$.
In terms of
\begin{eqnarray}
R(k;M) &\equiv& B(0;M,\infty,\infty) - B(k;M,\infty,\infty) 
\nonumber \\ 
&=&\left[ 4 M^2  +  4 \hat{k}^2 + (M^2 +  \hat{k}^2)^2\right]^{1/2} -
\left( 4 M^2 + M^4\right)^{1/2}\, , 
\end{eqnarray}
the large-$N$ saddle-point equation becomes
\begin{equation}
  \beta=\int_{-\pi}^{\pi} d_d k \,
\frac{1}{m + R(k;M)} \, .
\label{gapeq2}
\end{equation}
A continuous transition occurs when the defect length scale diverges, 
i.e. $m\rightarrow 0$.

Let us now consider the case $d=2$.  One can easily check that if
$M>0$ the integral (\ref{gapeq2}) diverges when $m\rightarrow 0$, as
expected since when $M>0$ the effective action represents a particular
regularization of the 2-$d$ nonlinear sigma model, which does not have
finite-temperature transitions according to the Mermin-Wagner theorem
\cite{MW-66}.  In this case, the magnetic susceptibility and 
the correlation length diverges in the limit
$\beta\rightarrow \infty$, as
\begin{equation} \label{nonlin}
\chi \sim \xi^2 
\sim e^{4 \pi C(M) \beta}, \qquad C(M)=\frac{2+M^2}{M \sqrt{4+M^2}}\, ,  
\end{equation}
which is the same behavior of the standard 2-$d$ O($N$) spin model with
$N>2$ and nearest-neighbor interactions, 
apart from a trivial normalization of the inverse temperature
$\beta$.  On the other hand, when $M=0$ a transition occurs at finite
temperature, since the integral (\ref{gapeq2}) is finite for
$m=0$. Indeed, one finds
\begin{equation}
\beta_c =\int_{-\pi}^{\pi} d_2 k \, \frac{1}{R(k;0)} = 0.252731 
\label{gapeq3}
\end{equation} 
and 
\begin{equation}
\chi \sim \xi \sim  t^{-1} \ln t^{-1}\, , 
\end{equation}
where $t\equiv (\beta_c-\beta)/\beta_c$.
Note that in this case the Mermin-Wagner theorem \cite{MW-66}
does not apply since long-range spin interactions are present. 

\begin{figure}[tb]
\centerline{\epsfig{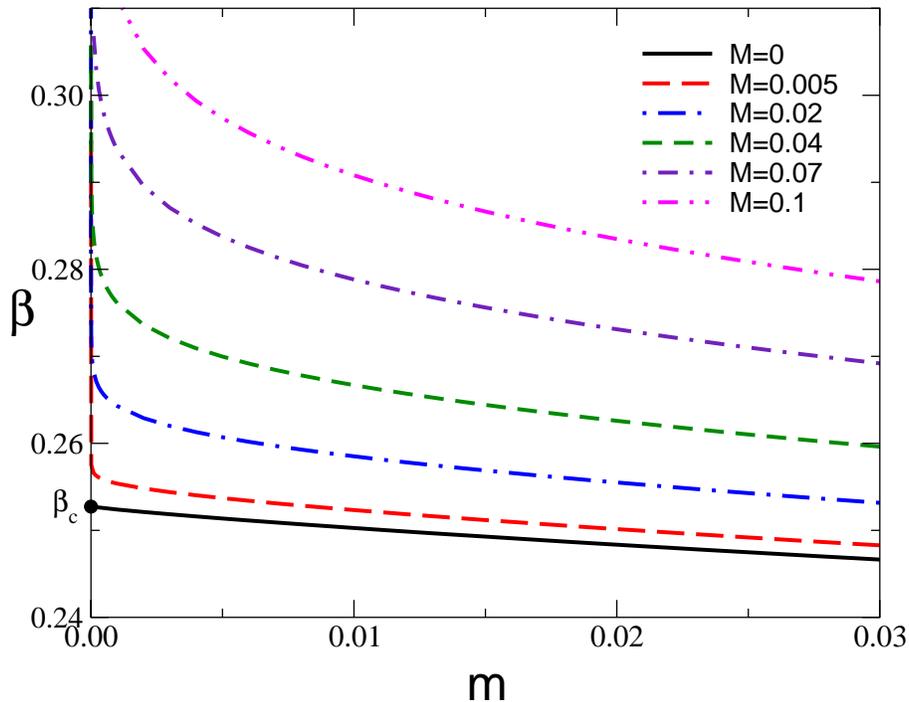}}
\vspace{2mm}
\caption{
The inverse temperature $\beta$ versus $m$ for several
values of $M$.}
\label{betafig}
\end{figure}

As already discussed, the critical point $T_c$ for $M=0$ is actually a
tricritical point.  Switching $M$ on, the critical behavior crossovers
to the more stable critical behavior for $M>0$ given by
eq.~(\ref{nonlin}).  This is shown in Fig.~\ref{betafig}, where the
function $\beta$ is plotted against $m$ for various values of $M$.
The curve for $M=0$ is approached by those for $M>0$ when
$M\rightarrow 0$, except in a small region $\Delta$ close to $m=0$, 
where they suddenly depart from the $M=0$ curve and diverge for
$m\rightarrow 0$. We note that $\Delta$ scales as $\Delta\sim M$
when $M\rightarrow 0$.
This is consistent with the expectation that the
crossover exponent $\phi$ is given by $\phi=1$, apart from logarithmic corrections.
We finally mention that the relation among the
quantities $\chi=1/(\beta m)$, $t$ and $M$ can be derived from equation
\begin{equation}
t \equiv {\beta_c - \beta\over \beta_c} = 1-\frac{1}{\beta_c}
\int_{-\pi}^{\pi} d_2 k \,
\frac{1}{m + R(k;M)} \, . 
\label{gapeq4}
\end{equation}

\section{Conclusions} 

In conclusion, we have shown how non-canonical quantum fields
$\varphi_i$ can be generated in $d$ dimensions using the interaction
of canonical fields $\Phi_i$ with a $\delta$-type 
(or more general dissipationless) defect in $d+1$
dimensions. The fields $\varphi_i$, which propagate on the defect,
define a unitary quantum field theory there. In order to clarify its
basic features, we analyzed the case in which an
$N$-component scalar field $\Phi_i$ in $d+1$ dimensions is
subject to a $\Phi^4$-type interaction localized on a $d$-dimensional 
defect. This theory can also be formulated in terms of
a $d$-dimensional O($N$) vector model, defined on the defect and
interacting with a $d+1$-dimensional bulk free field. 

General renormalization-group arguments and calculations in the large-$N$
limit allow to get a rather complete picture of the critical
behavior of the field theory induced on the
defect. The large-$N$ limit of the theory was studied within its
continuum formulation and also for statistical systems representing
a class of lattice regularizations. 
The main results can be summarized as follows.  When the bulk fields
$\Phi_i$ are massless, the induced $\varphi^4$ theory belongs to the
universality class of a specific statistical spin model with
long-range interactions, behaving as $\sim 1/r^{d+1}$.  The former
provides therefore a field theoretical description of the universal
critical properties of the latter.  The presence of a nonzero bulk
mass $M$ causes an interesting crossover to the critical behavior of
the standard O($N$) spin model with short-range interactions. We argue
that the critical point $T_{tc}$ at $M=0$ is actually a tricritical
point in the $T$-$M$-plane.  Following renormalization-group theory
applied to tricritical points, one can then consider a critical
crossover limit, defined when $t\sim T-T_{tc}\rightarrow 0$ and
$M\rightarrow 0$.  This presents a universal scaling behavior, which
can be studied within the continuous theory with non-vanishing bulk mass $M$. 

The generalization of this work to models involving gauge interactions 
opens interesting new possibilities and deserves further investigation.  

\appendix
\section{Some useful formulas}

In this appendix we provide a few details on the derivation of
eq.~(\ref{bbf}).  In order to determine the function $B$ from
eq.~(\ref{q11}), one needs to evaluate the matrix element 1,1 of the
inverse of a $n\times n$ matrix of the type
\begin{equation}
A_n = \left(
  \begin{array}{ccccc}
  a & -1 & 0 & \cdots & -1 \\
  -1& b &-1&\ddots& 0 \\
  0& \ddots& \ddots & \ddots & \vdots\\
  \vdots& & &  &-1 \\
  -1&0&\cdots &-1&b
  \end{array}
\right)\, .
\end{equation}
We are interested in the case $a>b\geq 2$.
This can be done by using the formula
\begin{equation}
[A_n^{-1}]_{1,1} = { {\rm det} A_n^{1,1}\over {\rm det} A_n }\, ,
\label{AN}
\end{equation}
where $A^{1,1}$ indicates the minor matrix corresponding to the 1,1
matrix element.  Let us introduce the $n\times n$ matrix
\begin{equation}
B_n = \left(
  \begin{array}{cccc}
   b&-1&0&\cdots \\
  -1&b &\ddots & \ddots \\
  0&\ddots &\ddots &-1 \\
  \vdots& \ddots &-1&b \end{array}
  \right)\, .
\end{equation}
Then the following relation holds
\begin{eqnarray}
&&{\rm det} A_n^{1,1} =  {\rm det} B_{n-1}, \nonumber \\
&&{\rm det} A_n =  a \, {\rm det} B_{n-1} - 2 \, {\rm det} B_{n-2} - 2\, .
\label{aeq}
\end{eqnarray} 
In order to compute the determinant of the matrix $B_n$, we note
that
\begin{equation}
{\det} B_1 = b\, , \qquad {\det} B_2 = b^2 - 1\, , 
\end{equation}
and the recursive formula
\begin{equation}
{\rm det} B_n = b \, {\rm det} B_{n-1} - {\rm det} B_{n-2} \, . 
\end{equation}
In the large-$n$ limit and for
$b>2$, ${\rm det} B_n$ diverges and 
\begin{equation}
\lim_{n\rightarrow \infty} 
{{\rm det} B_{n-1}\over  {\rm det} B_{n}}= 
{b-\sqrt{b^2-4} \over 2} < 1\, .
\end{equation}
This allows us to derive the formula
\begin{eqnarray}
\lim_{n\rightarrow \infty}  [A_n^{-1}]_{1,1} &=&
\lim_{n\rightarrow \infty}  
\left( a - 2 {{\rm det} B_{n-2}\over {\rm det} B_{n-1}} - 
{2 \over {\rm det} B_{n-1}} \right)^{-1} \nonumber \\
&=& 
{1\over a - b + \sqrt{b^2-4}}\, , 
\label{fineq}
\end{eqnarray}
which was used to obtain eq.~(\ref{bbf}).

\end{document}